\documentstyle[12pt]{article} 
\newcommand{\be}{\begin{equation}}
\newcommand{\nn}{\nonumber}
\newcommand{\bea}{\begin{eqnarray}}
\newcommand{\eea}{\end{eqnarray}}
\newcommand{\ba}{\begin{array}}
\newcommand{\ea}{\end{array}}
\newcommand{\ee}{\end{equation}}

\expandafter\ifx\csname mathbbm\endcsname\relax

\else

\fi \textheight 22cm \textwidth 15cm \topmargin 1mm \oddsidemargin
5mm \evensidemargin 5mm

\newcommand{\beas}{\begin{eqnarray*}}
\newcommand{\eeas}{\end{eqnarray*}}
\newcommand{\bes}{\begin{equation*}}
\newcommand{\ees}{\end{equation*}}

\newcommand{\lf}{\left}
\newcommand{\ri}{\right}
\newcommand{\f}{\frac}

\def\i2           {\mbox{$\frac{i}{2}$}}

\def\al           {\alpha}

\def\bet           {\beta}

\def\ft            {{\tilde F}_4}
\def\alt            {\tilde{\alpha}}
\def\bett            {\tilde{\beta}}
\def\lat            {\tilde{\la}}
\def\io            {{\rm Im}\, \Omega}

\def\del           {\delta}

\def\ep           {\epsilon}

\def\et           {\eta}

\def\ga           {\gamma}

\def\la           {\lambda}

\def\om           {\omega}
\def\omt            {\tilde \omega}
\def\et            {\tilde {e}}
\def\At             {\tilde {A}}

\def\ph           {\phi}

\def\si           {\sigma}
\def\Si           {\Sigma}

\def\th{\theta}

\def\eh  {{\hat e}}
\def\we {{\wedge}}

\def\eh    {{\hat{e}}} 
\begin{document}

\begin{titlepage}
\vspace*{20mm}
\begin{center}
{\LARGE \bf{{An $S^3\times S^3$ bundle over $S^4$ and new supergravity solutions}}}\\ 

\vspace*{1cm}
\vspace*{20mm} \vspace*{1mm} {Ali Imaanpur}

\let\thefootnote\relax\footnotetext{Email: aimaanpu@modares.ac.ir}

\vspace*{1cm}
  
{\it Department of Physics, School of Sciences\\ 
Tarbiat Modares University, P.O.Box 14155-4838, Tehran, Iran}

\vspace*{1mm}

\vspace*{1cm}

\end{center}

\begin{abstract}

The metric of $S^7$ can be written as an $SU(2)$-instanton bundle over $S^4$. It is also possible to write it differently as an anti-instanton bundle. We use this observation to construct an instanton--anti-instanton, $SU(2)\times SU(2)$, bundle over $S^4$. We show that this 10d manifold admits two Einstein metrics. We then rewrite the metric to isolate two $U(1)$ directions of the fibre and dimensionally reduce along them to get an eight dimensional metric describing an $S^2\times S^2$ bundle over $S^4$. This metric allows a harmonic 4-form which we use to derive new supergravity solutions in eleven and ten dimensions as $AdS_3\times M_8$ and $AdS_2\times M_8$, respectively.

\end{abstract}

\end{titlepage}

\section{Introduction} 
Among many different types of supergravity solutions, compactification to anti de Sitter (AdS) spaces are more typical, and depending on the compact manifold and the fluxes that one turns on different amounts of supersymmetries are preserved, or it is totally broken \cite{RUB, ENG, ROM, ENGTO, SOR}. Exact vacuum solutions, i.e., without source, are of particular interest in AdS/CFT duality, which relates string theory in AdS space to a particular gauge theory on its boundary. In this context, since supersymmetry puts strong constraints on the dynamics, supersymmetric solutions are more tractable on the gauge theory side. On the other hand, nonsupersymmetric solutions are closer to model realistic theories like QCD \cite{WIT}. 

$AdS_4\times S^7$ is an exact solution of eleven-dimensional supergravity which also shows up as the near horizon limit of an M2-brane solution. Writing the metric of $S^7$ as an $SU(2)$ bundle over $S^4$ has allowed to obtain new solutions where the fibre is squashed \cite{AWA, DUF, BAI, DUF3}. In fact, this way of expressing the metric reveals how the Yang-Mills solutions, as instantons on $S^4$, are embedded into the metric of $S^7$ as an Einstein manifold. Interestingly, one can replace the instanton  with an anti-instanton to get an equivalent metric on $S^7$. Now, from gauge theory we know that, if the gauge group is large enough, instanton--anti-instanton configurations are also a solution to the Yang-Mills equations, so it is curious to think if such bundles can be constructed on $S^4$. This is what we do in this paper by constructing an $SU(2)\times SU(2)$ bundle over $S^4$. The ten-dimensional metric is Einstein for two specific values of the squashing parameters. Noticing that the $S^7$ metric can also be written as a $U(1)$ bundle over ${\bf CP}^3$, we rewrite the ten-dimensional metric as a $U(1)\times U(1)$ bundle over an eight-dimensional base. It then becomes possible to reduce along these directions and get an eight-dimensional metric. This metric admits a harmonic 4-form which we then use to write down our ansatz and derive the solutions.    

In the next section we begin with recalling the $S^3$ fibration of $S^7$. The fibers are woven with an instanton connection over $S^4$. Replacing the instanton connection with an anti-instanton one gives an equivalent metric of $S^7$. We combine the two structures to write a 10-dimensional metric. The point of using an anti-instanton connection, instead of an instanton one, is to get the Ricci tensor with constant components, in the corresponding orthonormal basis. To derive supergravity solutions, in section 3, we rewrite the 10d metric so as to isolate two $U(1)$ fiber directions. This allows to reduce the 10d metric to an 8d metric simply by dropping the two $U(1)$ fibers, and at the same time have a Ricci tensor with constant components. We will show that the 8d manifold, $M_8$, endowed with this metric admits a harmonic 4-form, which we then use to write our ansatz for the 4-form field strength, $F_4$, in eleven-dimensional and type IIA supergravities. The solutions will be in the form of $AdS_3\times M_8$ and $AdS_2\times M_8$. Eleven-dimensional supergravity compactification to $AdS_3$, and type IIB compactification to $AdS_2$ have been studied earlier in \cite{PN, ALIB}.

\section{An $SU(2)\times SU(2)$ bundle over $S^4$}
In this section, we begin with describing the metric of $S^7$ as an $SU(2)$-instanton bundle over $S^4$, and then proceed to add an extra $SU(2)$-anti-instanton fiber to construct an $SU(2)\times SU(2)$ bundle over $S^4$. We could of course choose the extra fiber structure to be an instanton, but the corresponding Ricci tensor components of the metric would not be constant. The round metric on $S^7$ can be written as an $SU(2)$ bundle over $S^4$ \cite{AWA, DUF2},
\be
ds^2_{S^7}= d\mu^2 +\f{1}{4} \sin^2 \mu\, \Sigma_i^2 + (\sigma_i -\cos^2{\mu}/{2}\ \Sigma_i)^2\, ,\label{RO}
\ee
with $0\leq \mu \leq \pi$. $\Sigma_i$'s and $\sigma_i$'s are two sets of left-invariant one-forms 
\bea
&&\Si_1=\cos \ga\, d\al +\sin \ga \sin \al\,  d\bet\, , \nn \\
&&\Si_2=-\sin \ga\, d\al +\cos \ga \sin \al\,  d\bet\, , \  \nn \\
&& \Si_3= d\ga +\cos \al\,  d\bet \, , \nn
\eea
where $0\leq \ga \leq 4\pi ,\, 0\leq \al \leq \pi ,\, 0\leq \bet \leq 2\pi$, and 
with a similar expression for $\si_i$'s. They satisfy the $SU(2)$ algebra, namely,
\be
d\Sigma_i=-\f{1}{2}\, \ep_{ijk}\, \Sigma_j \we \Sigma_k\, , \ \ \ \  d\sigma_i=-\f{1}{2}\, \ep_{ijk}\, \sigma_j \we \sigma_k\, ,
\label{SIG}
\ee
with $i,j,k,\ldots =1,2,3$. 

Squashing corresponds to modifying the round metric on $S^7$ as follows
\be
ds^2_{S^7}= d\mu^2 +\f{1}{4} \sin^2 \mu\, \Sigma_i^2 + \la^2 (\sigma_i -A_i)^2\, ,\label{S7}
\ee
with $\la$ the squashing parameter. Here, to see explicitly how the instantons are embedded into the metric, we have written the fibers in terms of three $SU(2)$ one-form gauge fields
\be
A_i=\cos^2{\mu}/{2}\ \Sigma_i \, ,\label{A1}
\ee
where $i$ is considered as the algebra index. The gauge field strength is
\be
F_i=dA_i+\f{1}{2}\ep_{ijk} A^j\we A^k\, .
\ee
If we choose the following orthonormal basis of vielbeins 
\be
e^0=\, d\mu\, , \ \ e^i=\f{1}{2} \sin \mu\, \Sigma_i\, ,\ \ \ \ \eh^i=\la (\sigma_i - A_i)\, ,\label{BAS1}
\ee
we can see that the gauge field strengths of connection (\ref{A1}) are anti-self dual on the base manifold
\bea
F_1\!\!\!&=&\!\!\! dA_1+A_2\we A_3 = -e^0\we e^1 + e^2\we e^3\, ,\nn \\
F_2\!\!\!&=&\!\!\! dA_2+A_3\we A_1 = -e^0\we e^2 +e^3\we e^1\, ,\nn\\
F_3\!\!\!&=&\!\!\! dA_3+A_1\we A_2 = - e^0\we e^3 + e^1\we e^2\, .
\eea
Interestingly, we could have instead taken 
\be
\At_i=\sin^2{\mu}/{2}\ \Sigma_i \, ,\label{A2}
\ee
and get a self-dual field strength;
\bea
{\tilde F}_1\!\! &=&\!\! e^0\we e^1 + e^2\we e^3\, ,\nn \\
{\tilde F}_2\!\! &=&\!\! e^0\we e^2 +e^3\we e^1\, ,\nn \\
{\tilde F}_3\!\! &=&\!\! e^0\we e^3 + e^1\we e^2\, .
\eea

Either of gauge connections (\ref{A1}) or (\ref{A2}), could be used in (\ref{S7}) to get a metric with constant Ricci tensor components. In basis (\ref{BAS1}), we derive
\bea
R_{\al\bet}&=&\lf(3-\f{3\la^2}{2}\ri)\, \del_{\al\bet}\, , \nn \\
R_{\hat{\al}\hat{\bet}}&=& \lf(\f{1+2\la^4}{2\la^2}\ri)\, \del_{\hat{\al}\hat{\bet}}\, ,
\eea
where $\al, \bet =1,2,3,4$ and $\hat{\al}, \hat{\bet}=5,6,7$ are the base and the fiber indices, respectively. 
Note that the metric becomes Einstein for $\la =1$ and $\la=1/\sqrt{5}$.

Now, we would like to embed the above two structures into a single ten-dimensional metric, namely, an instanton--anti-instanton bundle over 
$S^4$. This is done simply by adding an extra 3-dimensional anti-instanton fiber to metric (\ref{S7});
\be
ds^2=  d\mu^2 +\f{1}{4} \sin^2 \mu\, \Sigma_i^2 + \la^2 
(\sigma_i -\cos^2 {\mu}/{2}\, \Sigma_i)^2+ \lat^2 (\tau_i -\sin^2 {\mu}/{2}\, \Sigma_i)^2\, ,\label{10MET}
\ee
with $\lat$ a constant parameter. Computing the Ricci tensor in the following orthonormal basis
\be
e^0=\, d\mu\, , \ \ e^i=\f{1}{2} \sin \mu\, \Sigma_i\, ,\ \ \ \ \eh^i=\la (\sigma_i - A_i)\, ,\ \ \ \ \et^i=\la (\tau_i - \At_i)\, , \label{BAS2}
\ee
results in
\bea
R_{\al\bet}&=&\lf(3-\f{3\la^2}{2}-\f{3\lat^2}{2}\ri)\, \del_{\al\bet}\, , \nn \\
R_{\hat{\al}\hat{\bet}}&=& \lf(\f{1+2\la^4}{2\la^2}\ri)\, \del_{\hat{\al}\hat{\bet}}\, ,\nn\\
R_{{\alt}{\bett}}&=& \lf(\f{1+2\lat^4}{2\lat^2}\ri)\, \del_{{\alt}{\bett}}\, ,
\eea
where ${\alt}, {\bett}=8,9,10$ are the tangent indices of the extra anti-instanton fiber. We observe that for $\la^2=\lat^2=1/2$, and  $\la^2=\lat^2=1/4$, the metric is Einstein, with $R_{MN}=3/2\, \del_{MN}$, and $R_{MN}=9/4\, \del_{MN}$, respectively, ($M$ and $N$ being the collective indices). In the present method of deriving supergravity solutions it is crucial for the Ricci tensor components to be constant. Had we used instanton connections as an extra 3-dimensional fiber, the Ricci tensor would not have had constant components. 

It is interesting to examine the structure of the metric a bit further, and show that it indeed admits a harmonic 3-form. Let us first introduce $\om_3$ and $\omt_3$, the volume elements of the instanton and the anti-instanton fibers, respectively:
\be
\om_3 = \eh^1\we \eh^2 \we \eh^3\, ,\ \ \ \ \omt_3= \et^1\we \et^2 \we \et^3\, ,
\ee
taking the exterior derivative, we obtain
\be
d\om_3 = \f{\la}{2}\lf( \ep_{ijk}\, e^0\we e^i \we \eh^j \we \eh^k +\, e^i\we e^j\we \eh^i\we \eh ^j\ri)\, .\label{OM3}
\ee
The Hodge dual, excluding the extra three anti-instanton fibers, reads
\be
*_7d\om_3={\la}\, \eh^i\we\, (e^0\we e^i +\f{1}{2}\ep_{ijk}\, e^j\we e^k )\, ,
\ee
so, we derive
\be
d*_7d\om_3 ={6\la^2}\, \om_4 - \f{1}{\la}\, d\om_3 \, ,\label{STA}
\ee
where
\be
\om_4= e^0\we e^1\we e^2\we e^3 \, ,
\ee
is the volume element of the base. Note that $\om_4$ is closed; $d\om_4 =0$. Similarly, 
\be
d{\tilde *}_7d\omt_3 =-{6\lat^2}\, \om_4 - \f{1}{\lat}\, d\omt_3 \, ,\label{STA2}
\ee
with
\be
{\tilde *}_7d\omt_3={\lat}\, \et^i\we\, (e^0\we e^i -\f{1}{2}\ep_{ijk}\, e^j\we e^k )\, .
\ee
As with $*_7$ which defines the Hodge dual with anti-instanton fibers excluded, ${\tilde *}_7$ is the Hodge dual with instanton fibers excluded. Now, we claim that the metric admits a harmonic 3-form. For simplicity, let us set $\la=\lat$. Looking at eqs. (\ref{STA}) and (\ref{STA2}), we observe that
\be
\Omega_3= *_7d\om_3 +{\tilde *}_7d\omt_3 +\f{1}{\la}(\om_3 +\omt_3)\, ,
\ee
is closed. Taking the Hodge dual, we get
\be
*\Omega_3= d\om_3\we \omt_3 +\om_3 \we d\omt_3 +\f{1}{\la}\om_4\we (\omt_3 -\om_3)\, ,
\ee
this is also closed since $d\om_3\we d\omt_3=0$, as $d\om_3$ is self-dual (with respect to the base manifold $S^4$) while $d\omt_3$ is anti-self-dual. Also note that $\om_4\we d\om_3=\om_4\we d\omt_3=0$. So, we conclude that
\be
d\Omega_3=d*\Omega_3=0\, ,
\ee
namely, the 10-dimensional manifold with metric (\ref{10MET}) admits a harmonic 3-from. 

Using the three closed 4-forms, $\om_4, d\om_3$, and $d\omt_3$, we could proceed along the lines of \cite{ALI} to try an ansatz for the 4-form field, $F_4$, and solve the equations of motion. However, in type IIA case, the number of independent equations will exceed the number of free parameters, and so there will be no solution. For eleven-dimensional supergravity, we may consider a trivial circle bundle, $S^1\times M_{10}$, with $M_{10}$ the 10d manifold we constructed in this section. But, in this case, the energy-momentum tensor of $F_4$ backreacts on $S^1$, and since $S^1$ is flat there will be no solution. One way out is to observe that the 10d metric can be rewritten as a $U(1)\times U(1)$ bundle over an 8d manifold. The $U(1)$ factors can then be scaled away leaving an eight-dimensional manifold. This is what we do in the next section.

\section{$S^2\times S^2$ bundle over $S^4$}
In the beginning of the previous section, the metric of $S^7$ was written as an $SU(2)$ bundle over $S^4$. It is also possible to write the 
metric as a $U(1)$ bundle over ${\bf CP}^3$. Further, it is observed that ${\bf CP}^3$ itself can be written as an $S^2$ bundle over $S^4$. In this form one can construct a family of  homogeneous metrics by rescaling the fibers \cite{HOPF, FONT}. In this section, first we rewrite the $S^7$ metric, (\ref{RO}), as a $U(1)$ bundle over ${\bf CP}^3$ \cite{ALI}, and then repeat the same process for the 10d metric we constructed in the previous section. This allows to isolate two $U(1)$ directions of the metric and then  dimensionally reduce along them to get an 8-dimensional metric. 

Beginning with metric (\ref{S7}), note that we can also write it as 
\bea
ds^2_{S^7}\!&=&\! d\mu^2 +\f{1}{4} \sin^2 \mu\, \Sigma_i^2 +\la^2(\sigma_i -\cos^2{\mu}/{2}\ \Sigma_i)^2\nn \\
\!&=&\! d\mu^2 +\f{1}{4} \sin^2 \mu\, \Sigma_i^2 + \la^2 \sin^2 \th_1\, (d\ph_1 -\cot \th_1 (\cos \ph_1 A_1 +\sin \ph_1 A_2)+A_3)^2 \nn \\
\!&+&\!\la^2 (d\th_1 -\sin \ph_1 A_1+\cos \ph_1 A_2)^2+\la^2(d\tau-A)^2\, , \label{NEE}
\eea 
where
\be
A= \cos \th_1 \, d\ph_1+\sin \th_1 (\cos \ph_1 A_1 +\sin \ph_1 A_2)+\cos \th_1 A_3 \, , \label{GAU}
\ee
and $\sigma_i$'s are left-invariant one-forms that are chosen as follows: 
\bea
&&\si_1= \sin \ph_1\, d\th_1 +\sin \th_1 \cos \ph_1 \,  d\chi_1\, ,\nn \\
&&\si_2= -\cos \ph_1\, d\th_1 +\sin \th_1 \sin \ph_1\,  d\chi_1\, , \  \nn \\
&& \si_3= -d\ph_1 +\cos \th_1\,  d\chi_1 \, . \nn
\eea
The point of writing the $S^7$ metric in this form is that the last $U(1)$ factor in (\ref{NEE}) can further be rescaled, i.e., $\la \to \la'$, so that the Ricci tensor components (in a basis we introduce shortly) are still constant. We use this observation to rewrite 10d metric (\ref{10MET}) as a $U(1)\times U(1)$ bundle over a base, which itself is an $S^2\times S^2$ bundle over $S^4$: 
\bea
ds^2_{10}\!\! &=&\!\! d\mu^2 +\f{1}{4} \sin^2 \mu\, \Sigma_i^2 +\la^2 (d\th_1 -\sin \ph_1 A_1+\cos \ph_1 A_2)^2 \nn \\
\!\! &+&\!\! \la^2 \sin^2 \th_1\, (d\ph_1 -\cot \th_1 (\cos \ph_1 A_1 +\sin \ph_1 A_2)+A_3)^2 +\la'^2(d\chi_1-A)^2\nn \\
\!\! &+&\!\! \lat^2 (d\th_2 -\sin \ph_2 \At_1+\cos \ph_2 \At_2)^2 \nn \\
\!\! &+&\!\! \lat^2 \sin^2 \th_2\, (d\ph_2 -\cot \th_2 (\cos \ph_2 \At_1 +\sin \ph_2 \At_2)+\At_3)^2 +\lat'^2(d\chi_2-\At)^2 \label{NE0}
\eea 
whereas, just as in (\ref{GAU}), $\At$ is defined in terms of $\At_i$;
\be
\At= \cos \th_2 \, d\ph_2+\sin \th_2 (\cos \ph_2 \At_1 +\sin \ph_2 \At_2)+\cos \th_2 \At_3 \, . \label{GAU2}
\ee
Note that here we have taken $\tau_i$'s in (\ref{10MET}) as follows: 
\bea
&&\tau_1= \sin \ph_2\, d\th_2 +\sin \th_2 \cos \ph_2 \,  d\chi_2\, ,\nn \\
&&\tau_2= -\cos \ph_2\, d\th_2 +\sin \th_2 \sin \ph_2\,  d\chi_2\, , \  \nn \\
&& \tau_3= -d\ph_2 +\cos \th_2\,  d\chi_2 \, , \nn
\eea
and rescaled two $U(1)$ fibers in (\ref{NE0}) as $\la\to\la' \, ,\ \lat\to \lat'$. 

Choosing the following basis
\bea
&& e^0=\, d\mu\, , \ \ \ \ e^i=\f{1}{2} \sin \mu\, \Sigma_i\, ,\nn \\ 
&& e^5=\la (d\th_1 -\sin \ph_1 A_1+\cos \ph_1 A_2) \, ,\nn \\
&& e^6=\la \sin \th_1 (d\ph_1 -\cot \th_1 (\cos \ph_1 A_1 +\sin \ph_1 A_2)+A_3)\, ,\nn \\
&& e^7=\la' (d\chi_1 -A)\, , \nn \\
&& \et^5=\lat (d\th_2 -\sin \ph_2 \At_1+\cos \ph_2 \At_2) \, ,\nn \\
&& \et^6=\lat \sin \th_2 (d\ph_2 -\cot \th_2 (\cos \ph_2 \At_1 +\sin \ph_2 \At_2)+\At_3)\, ,\nn \\
&& \et^{7}=\lat' (d\chi_2 -\At)\, , \label{VI}
\eea
the Ricci tensor turns out to be diagonal and reads
\bea
&& R_{00}=R_{11}=R_{22}=R_{33}=3-\la^2-\lat^2-{\la'^2}/{2}-{\lat'^2}/{2} \, ,\nn \\
&& R_{55}=R_{66}=\la^2+{1}/{\la^2}-{\la'^2}/{2\la^4}\, ,\ \ \ \ \ R_{77}=\la'^2+{\la'^2}/{2\la^4}\, , \nn \\
&& R_{88}=R_{99}=\lat^2+{1}/{\lat^2}-{\lat'^2}/{2\lat^4}\, ,\ \ \ \ \ R_{10,10}=\lat'^2+{\lat'^2}/{2\lat^4}\, .\label{EE}
\eea 
This explicitly shows that we can rescale two $U(1)$ fiber directions along $\chi_1$ and $\chi_2$ and still have a diagonal Ricci tensor with constant components. In fact, we can set $\la'=\lat'=0$ in (\ref{NE0}) and get an 8-dimensional metric. In the following, we show that this manifold admits a harmonic 4-form which is then used to derive a supergravity solution in 11 dimensions.  

\subsection{The ansatz for $F_4$}
To discuss our ansatz for $F_4$ and the solutions, henceforth, we take the following eight-dimensional metric
\bea
ds^2_{8}&=& d\mu^2 +\f{1}{4} \sin^2 \mu\, \Sigma_i^2 +\la^2 (d\th_1 -\sin \ph_1 A_1+\cos \ph_1 A_2)^2 \nn \\
&+&\la^2 \sin^2 \th_1\, (d\ph_1 -\cot \th_1 (\cos \ph_1 A_1 +\sin \ph_1 A_2)+A_3)^2  \nn \\
&+&\lat^2 (d\th_2 -\sin \ph_2 \At_1+\cos \ph_2 \At_2)^2 \nn \\
&+&\lat^2 \sin^2 \th_2\, (d\ph_2 -\cot \th_2 (\cos \ph_2 \At_1 +\sin \ph_2 \At_2)+\At_3)^2\, ,  \label{NE}
\eea 
and assume that the eleven-dimensional manifold is given by $AdS_3\times M_8$, with the metric
\be
ds_{11}^2 = ds_{3}^2 + ds_{8}^2\, . 
\ee
As in \cite{ALI}, it also proves useful to define the following self-dual 2-forms
\bea
R_1\!\!&=&\!\! \sin \ph_1 (e^{01}+e^{23}) -\cos \ph_1(e^{02}+e^{31}) \, ,\nn \\
R_2\!\!&=&\!\! \cos \th_1\cos \ph_1 (e^{01}+e^{23})+ \cos \th_1 \sin \ph_1 (e^{02}+e^{31})-\sin \th_1 (e^{03}+e^{12})\, ,\nn \\
K\!\!&=&\!\!  \sin \th_1\cos \ph_1 (e^{01}+e^{23})+ \sin \th_1 \sin \ph_1 (e^{02}+e^{31})+\cos \th_1 (e^{03}+e^{12})\, ,\label{THREE}
\eea
where we have used the shorthand notation $e^{ij}=e^i\we e^j$. The key feature of this definition, that we will use frequently in this paper, is that these three forms are orthogonal to each other, i.e.,
\be
R_1\we R_2=K\we R_1=K\we R_2=0\, .\label{WED}
\ee
We can also define a set of anti self-dual 2-forms which are orthogonal (with respect to the wedge product) to the above self-dual set:
\bea
{\tilde R}_1\!\!&=&\!\! \sin \ph_2 (-e^{01}+e^{23}) -\cos \ph_2(-e^{02}+e^{31}) \, ,\nn \\
{\tilde R}_2\!\!&=&\!\! \cos \th_2\cos \ph_2 (-e^{01}+e^{23})+ \cos \th_2 \sin \ph_2 (-e^{02}+e^{31})-\sin \th_2 (-e^{03}+e^{12})\, ,\nn \\
{\tilde K}\!\!&=&\!\!  \sin \th_2\cos \ph_2 (-e^{01}+e^{23})+ \sin \th_2 \sin \ph_2 (-e^{02}+e^{31})+\cos \th_2 (-e^{03}+e^{12}) \label{THREE2}
\eea
and obey relations like the ones in  (\ref{WED}).  $K$ and ${\tilde K}$ naturally appear in the field strength of $A$ and $\At$, respectively;
\bea
F\!\!&=&\!\!dA =- K -e^{56}/{\la^2}\, ,\nn \\
{\tilde F}\!\!&=&\!\!d\At =- {\tilde K} -\et^{56}/{\lat^2}\, .\label{F} 
\eea
Taking the exterior derivative, it is easy to see that
\bea
dK\!\!&=&\!\! - \f{1}{\la^2}de^{56}=-\f{1}{\la}\, \io\, ,\nn \\
d{\tilde K}\!\!&=&\!\! - \f{1}{\lat^2}d\et^{56}=-\f{1}{\lat}\, {{\rm Im}\, {\tilde \Omega}}\, ,\label{2K}
\eea
where,
\bea
{\rm Im}\, \Omega =R_1\we e^6 - R_2 \we e^5\, ,\nn \\
{{\rm Im}\, {\tilde \Omega}}={\tilde R}_1\we \et^6 - {\tilde R}_2 \we \et^5\, .
\eea
Here, $\Omega$ is the holomorphic 3-form of the ${\bf CP}^3$ which appears in the metric of $S^7$ viewed as a $U(1)$ bundle over ${\bf CP}^3$, and it should not be confused with $\Omega_3$ of the previous section, and $\Omega_4$ that will appear in the next section. From (\ref{2K}), it trivially follows that $d\io= d {{\rm Im}\,{\tilde \Omega}}=0$. 

Now, let us examine if there is any harmonic 4-form on the 8d manifold endowed with metric (\ref{NE}). A closed 4-form which is easy to guess is
\be
\Omega_4=\al \om_4 +\bet K\we e^{56} + \ga {\tilde K}\we \et^{56}\, ,
\ee
for constant coefficients $\al, \bet, \ga$. The first term is closed, $d\om_4=0$. Using (\ref{2K}) together with $\io \we e^{56}= \io \we K=0$, and  $ {{\rm Im}\,{\tilde \Omega}} \we \et^{56}=  {{\rm Im}\,{\tilde \Omega}} \we {\tilde K}=0$, we see that the second and the third terms are also separately closed, so $d\Omega_4=0$. For the Hodge dual we have
\be
*_8\Omega_4=\al e^{56}\we \et^{56}+\bet K\we \et^{56} -\ga {\tilde K}\we e^{56}\, ,
\ee
the minus sign is because ${\tilde K}$ is anti self-dual. Demanding $d*_8\Omega_4=0$, and noticing that 
\be
{{\rm Im}\,{\tilde \Omega}} \we K=0\, , \ \ \ \  \io \we  {\tilde K}=0\, ,
\ee
as untilted and tilted 2-forms are self-dual and anti self-dual, respectively, the $\bet$ and $\ga$ get fixed in terms of an overall constant $\al$. So, here is the harmonic 4-form;
\be
\Omega_4= \om_4 +\la^2 K\we e^{56} -\lat^2 {\tilde K}\we \et^{56}\, .
\ee
Either $\Omega_4$ or $*_8\Omega_4$ can be used as an ansatz for the 4-form field strength in its 11d supergravity equation of motion
\be
d*_{11}F_4=-\f{1}{2}\, F_4\we\, F_4 \, .\label{MAX}
\ee
Let us take
\be
F_4=\al *_8\Omega_4= \al (e^{56}\we \et^{56} +\la^2 K\we \et^{56} +\lat^2 {\tilde K}\we e^{56})\, ,\label{ANSS}
\ee
since $*\Omega_4\we *\Omega_4=0$, the right hand side of (\ref{MAX}) becomes zero. Therefore, since $*\Omega_4$ is harmonic, $F_4$ satisfies both the Bianchi identity and the equation of motion, (\ref{MAX}). One might think that similar ansatz may also work for simpler 8d manifolds like $S^4\times S^4$ or ${\bf CP}^4$, which admit 4-form harmonics too. The point, however, is that for such ansatz the right hand side of Maxwell equation does not vanish. So, here, it is important that we have  $*\Omega_4\we *\Omega_4=0$. Having found a solution for the Maxwell equation, (\ref{MAX}), let us now turn to the eleven-dimensional Einstein equations;
\be
R_{MN}=\f{1}{12}\, F_{MPQR}F_N^{\, \ PQR} -\f{1}{3\cdot 48}\, g_{MN}\ F_{PQRS}F^{PQRS}\, ,
\ee
where $M,N,P,\ldots =0,1,\ldots, 10$. Setting for simplicity $\la=\lat$, the Ricci tensor components of $M_8$ on the left hand side can be read off from (\ref{EE}) (with $\la'=\lat'=0$). Further, with ansatz (\ref{ANSS}), we can calculate the right hand side of the above equations to get;
\bea 
3-2\la^2 =\al^2\lf(\f{\la^4}{3} -\f{1}{6}\ri)\, , \nn \\
\f{1+\la^4}{\la^2}= \al^2 \lf(\f{\la^4}{3} +\f{1}{3}\ri)\, .\label{ALG}
\eea
The solution reads
\be
\la^2=\f{1}{6}(\sqrt{ 15}+3)\, , \ \ \ \ \al^2=3(\sqrt {15}-3)\, .
\ee
As $\la^2>1$, the $S^2\times S^2$ fibers in (\ref{NE}) are stretched with respect to the base. For the Ricci tensor along $AdS_3$, we obtain
\be
R_{\mu\nu}= -\f{1}{6}\lf(5\sqrt{ 15}-3\ri)g_{\mu\nu}\, ,
\ee
with $\mu,\nu =0,1,2$. 

If we take $F_4=\al \Omega_4$, it is clear that Maxwell equation (\ref{MAX}) is satisfied. However, for this ansatz we have found no real solution of the Einstein equations. 

\section{10d IIA supergravity solution}
The ansatz we found in the previous section can also be used to obtain a ten-dimensional type IIA supergravity solution. For type IIA field equations in the string frame we have
\bea
&& d\ft=-F_2\wedge H \, , \ \ \ \ d*\ft=-\ft\wedge H \, , \ \ \ \ dH=0\, , \ \ \ \ dF_2=0\, ,\nn \\ 
&& d*(e^{-2\ph}H)=-F_2\wedge *\ft+\f{1}{2}\ft\wedge \ft \, , \ \ \ \ d*F_2=H\wedge *\ft \, , \nn \\
&& d*d\ph-d\ph\wedge *d\ph -\f{1}{8}H\wedge *H +\f{1}{4\cdot 3!}R\, \ep_4\we J^3=0\, ,\nn \\
&& \ft=F_4-A_1\wedge H \, .\label{FEQ}
\eea
If we only turn on $F_4=\al *\Omega_4$ as in the previous section, since $F_4\we F_4=0$ we can set $H=F_2=0$. But then $R$, the scalar curvature, will not be zero and hence, acts via the last equation as a source for the dilaton, $\phi$. However, if we further turn on $F_2=\eta \ep_2$, with $\ep_2$ the volume element of an $AdS_2$ factor, we can satisfy $R=0$, and have a constant dilaton. Our ansatz for the metric thus will be
\be
ds_{10}^2=ds_{AdS_2}^2 +ds_8^2\, ,
\ee
together with
\be
F_4=\al *\Omega_4\, , \ \ \ \ F_2=\eta\ep_2\, ,\ \ \ \ H=0\, , \ \ \ \ \phi=const.\label{ANS3}
\ee
We can check it easily that with this ansatz all the equations of motion, (\ref{FEQ}), are satisfied. For the Einstein equations we have,
\bea
R_{MN}&=&\f{e^{2\ph}}{12}\, \lf(F_{MPQR}F_N^{\, \ PQR} -\f{3}{32}\, g_{MN}\ F_{PQRS}F^{PQRS}\ri)\,  \nn \\
&+&\f{e^{2\ph}}{2}\, \lf(F_{MP}F_N^{\, \ P} -\f{1}{4}\, g_{MN}\ F_{PQ}F^{PQ}\ri)\, ,
\eea
so requiring $R=0$, fixes the relative scale of $F_4$ and $F_2$;
\be
\eta^2=\f{\al^2}{3}(1+4\la^4)\, .
\ee
Using (\ref{EE}) and ansatz (\ref{ANS3}), the Einstein equations along $M_8$ become
\bea 
3-2\la^2 =e^{2\phi}\al^2\lf(\f{\la^4}{3} -\f{1}{6}\ri)\, , \nn \\
\f{1+\la^4}{\la^2}= e^{2\phi}\al^2 \lf(\f{\la^4}{3} +\f{1}{3}\ri)\, ,
\eea
which are the same equations as in (\ref{ALG}), with the solution
\be
\la^2=\f{1}{6}(\sqrt{ 15}+3)\, , \ \ \ \ e^{2\phi}\al^2=3(\sqrt {15}-3)\, .
\ee
For the Ricci tensor along $AdS_2$ we obtain
\be
R_{\mu\nu} =-\f{1}{3}\lf(5\sqrt{15}-3\ri)\,  g_{\mu\nu}\, ,
\ee
with $\mu,\nu =0,1$. 

The above type IIA solution can be understood as a reduction of the  $AdS_3\times M_8$ solution of the previous section. In fact, $AdS_3$ metric can be written as a $U(1)$ bundle over $AdS_2$. If we dimensionally reduces along this $U(1)$ direction, then, in ten dimensions, $F_2$ appears as the field strength of the corresponding $U(1)$ connection. 

\section{Conclusions}
In this paper, we studied adding an extra structure of anti-instanton fibers to the metric of $S^7$, and constructed a 10d metric, which had constant Ricci tensor components in the corresponding orthonormal basis. The metric also admitted a harmonic 3-form. To get a supergravity solution, however, we argued that the metric could be rewritten as a $U(1)\times U(1)$ bundle over an 8d manifold. In this way, it was possible to scale away the $U(1)$ directions and get a new 8d manifold with metric (\ref{NE}). We showed that this manifold admits a harmonic 4-form, which we then used as an ansatz for the 4-form field strength in eleven-dimensional and ten-dimensional type IIA supergravities. At the end, we saw that, through this choice of the metric and the ansatz for $F_4$, the equations of motion could be reduced to a set of simple algebraic equations. The solutions, in turn, fixed the free parameter $\la$, the scale factor of fibers in the metric, and the overall scale of $F_4$ (and $F_2$ in the case of type IIA).

\hspace{30mm}


\vspace{1.5mm}

\noindent

\vspace{1.5mm}

\noindent

\newpage



\end{document}